\documentclass{article}
\usepackage{spconf,amsmath,graphicx,hyperref}
\usepackage{multirow}
\usepackage{multicol}
\usepackage{colortbl}
\usepackage{makecell}
\usepackage{amssymb}

\title{Perceptual Quality Optimization of Image Super-Resolution}

\name{Wei Zhou$^{1}$, Yixiao Li$^{1}$, Hadi Amirpour $^{2}$, Xiaoshuai Hao$^{3}$, Jiang Liu$^{1}$, Peng Wang$^{4}$, Hantao Liu$^{1}$\thanks{Corresponding author: Wei Zhou (zhouw26@cardiff.ac.uk). Co-first author: Yixiao Li.}}
\address{%
$^{1}$ School of Computer Science and Informatics, Cardiff University, UK
\\
$^{2}$ Christian Doppler Laboratory ATHENA, Alpen-Adria-Universität, Klagenfurt, Austria
\\
$^{3}$ XiaomiEV, China
$^{4}$ Centre for Vision, Speech and Signal Processing, University of Surrey, UK
}
\begin{document}
%
\maketitle
\begin{abstract}
Single image super-resolution (SR) has achieved remarkable progress with deep learning, yet most approaches rely on distortion-oriented losses or heuristic perceptual priors, which often lead to a trade-off between fidelity and visual quality. To address this issue, we propose an \textit{Efficient Perceptual Bi-directional Attention Network (Efficient-PBAN)} that explicitly optimizes SR towards human-preferred quality. Unlike patch-based quality models, Efficient-PBAN avoids extensive patch sampling and enables efficient image-level perception. The proposed framework is trained on our self-constructed SR quality dataset that covers a wide range of state-of-the-art SR methods with corresponding human opinion scores. Using this dataset, Efficient-PBAN learns to predict perceptual quality in a way that correlates strongly with subjective judgments. The learned metric is further integrated into SR training as a differentiable perceptual loss, enabling closed-loop alignment between reconstruction and perceptual assessment. Extensive experiments demonstrate that our approach delivers superior perceptual quality. Code is publicly available at \url{https://github.com/Lighting-YXLI/Efficient-PBAN}.
\end{abstract}
\begin{keywords}
Image super-resolution, quality assessment, bi-directional attention, perceptual optimization
\end{keywords}
\section{Introduction}
\label{sec:intro}

Single image super-resolution (SR) aims to reconstruct a high-resolution (HR) image from its low-resolution (LR) counterpart, and has been a long-standing problem in image processing and computer vision. With the success of deep learning, SR methods based on convolutional neural networks \cite{dong2015image,zhang2018image} and transformer architectures \cite{liang2021swinir} have achieved remarkable improvements in terms of distortion-oriented metrics such as peak signal-to-noise ratio (PSNR) and structural similarity index (SSIM) \cite{SSIM}. Such models excel at signal fidelity but often fail to recover high-frequency details critical for human perception. Consequently, optimizing solely for these fidelity measures frequently leads to perceptually unsatisfying results, such as over-smoothed textures and unnatural appearances.  

To enhance perceptual quality, researchers have proposed perceptual losses that leverage features from pre-trained classification networks \cite{johnson2016perceptual} and adversarial learning frameworks such as SRGAN \cite{ledig2017photo}. Although adversarial training improves realism, the generated textures are often unstable or contain hallucinations. More recently, diffusion-based generative models have been applied to SR \cite{saharia2022image, li2022srdiff, wang2024exploiting}, showing impressive perceptual realism by iteratively refining image details. While achieving state-of-the-art visual quality, such methods usually suffer from heavy computation and long inference time, limiting their practicality in real-world applications.  

In parallel, the field of image quality assessment (IQA) has advanced rapidly, with deep learning-based metrics trained on human opinion scores demonstrating strong correlation with perceptual judgments \cite{bosse2017deep, ying2020patches, ke2021musiq,wang2024blind}. Most existing IQA metrics, however, are trained on generic distortions such as noise or blur, which do not accurately reflect the characteristic artifacts produced by SR algorithms. Therefore, existing databases are often generic and do not adequately cover recent SR methods, making them less effective for perceptual optimization in SR.  

\begin{figure*}[t]
  \centering
  \vspace{-2em}
  \includegraphics[width=0.95\textwidth]{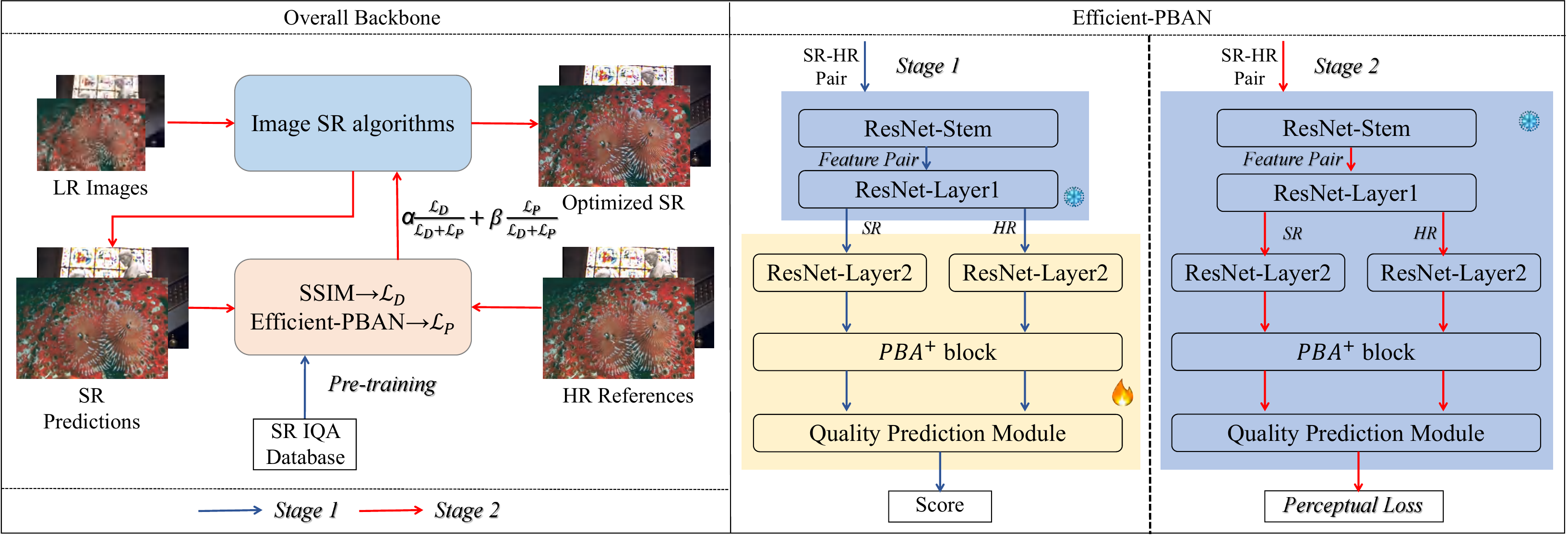}
  \caption{Overall framework of the proposed method. Left: SR optimization pipeline driven by Efficient-PBAN. Right: Efficient-PBAN structure with two-stage training, where the “ice" and “fire" icons denote frozen and trainable parameters, respectively. And the details of PBA$^{+}$ block are described in Section \ref{section21}. $\mathcal{L}_{D}$ and $\mathcal{L}_{P}$ refer to the distortion-oriented and perceptual loss, respectively.}
  \label{fig:framework}

\end{figure*}

To bridge this gap, we construct the latest SR quality database~\cite{li2025vquala2025challengeimage}. This built database enables the training of a perceptual metric that is customized for SR. However, existing SR metrics (e.g., PFIQA~\cite{pfiqa}, PBAN~\cite{li2025pban}) are patch-based, requiring extensive patch sampling and calculation, which greatly limits their applicability as end-to-end loss functions. Building upon these foundations, we propose an \textit{Efficient Perceptual Bi-directional Attention Network (Efficient-PBAN)} for the perceptual quality optimization of image SR. Efficient-PBAN is first trained on the SR quality database to predict perceptual quality aligned with human judgments, and then integrated into SR optimization as a differentiable perceptual loss. This closed-loop design directly aligns SR reconstruction with perceptual assessment. Experimental results on standard benchmarks confirm its effectiveness in enhancing perceptual quality, while still maintaining competitive distortion fidelity. These together establish Efficient-PBAN as a practical paradigm for perceptual quality optimization in SR.

The main contributions of this work are summarized as follows:  
\begin{itemize}
    \item We introduce the latest SR quality database that covers a wide range of recent SR methods with human quality ratings, providing a solid foundation for perceptual metric learning.  
    \item We propose Efficient-PBAN, a bi-directional attention quality prediction network that achieves strong correlation with subjective judgments while being lightweight and efficient.  
    \item We integrate the learned metric into SR optimization as a differentiable perceptual loss, achieving superior perceptual quality on SR benchmarks.  
\end{itemize}
\begin{table*}[t]
\centering
\small
\caption{Quantitative comparison based on CAMixerSR~\cite{wang2024camixersr} and LINF~\cite{linf} across B100~\cite{B100} and DIV2K~\cite{DIV2K} databases with different optimization counterparts. Results are reported as mean score on each testing database. Best and second-best results for each database and baseline are highlighted in \textcolor{red}{red} and \textcolor{blue}{blue}, respectively. Note that EPBAN refers to the Efficient-PBAN.}
\begin{tabular}{c|c|c|cccccccc}
\Xhline{1pt}
Test               & SR                   & Model                                                           & PSNR $\uparrow$                                                        & SSIM $\uparrow$                                                       & LPIPS $\downarrow$                                                      & PFIQA  $\uparrow$                                                     & EPBAN $\uparrow$                                             & SF $\uparrow$                                                         & SN$\uparrow$                                                          & SFSN$\uparrow$                                                        \\ \Xhline{1pt}
\multirow{4}{*}{\cite{B100}}  & \multirow{8}{*}{~\cite{wang2024camixersr}} & Original                                                        & \textcolor{red}{26.3513} & \textcolor{blue}{0.7188} & \textcolor{blue}{0.3382} & 3.8528 & 2.8682 & \textcolor{red}{0.9857} & 2.1952 & 1.1067 \\ \cline{3-11} 
                       &                            & +SSIM                                                           & \textcolor{blue}{25.6755} & \textcolor{red}{0.7304} & \textcolor{red}{0.3358} & 3.8702 & 2.9415 & 0.9764 & 2.3886 & 1.1177 \\ \cline{3-11} 
                       &                            & +EPBAN                                                 & 25.6203 & 0.6963 & 0.3567 & \textcolor{blue}{3.8911} & \textcolor{blue}{3.3165} & 0.9836 & \textcolor{red}{2.4982} & \textcolor{red}{1.1351} \\ \cline{3-11} 
                       &                            & \begin{tabular}[c]{@{}c@{}}+SSIM$\&$EPBAN\end{tabular} & 25.6741 & 0.7110 & 0.3560 & \textcolor{red}{3.9348} & \textcolor{red}{3.4025} & \textcolor{blue}{0.9843} & \textcolor{blue}{2.4084} & \textcolor{blue}{1.1268} \\ \Xcline{1-1}{1pt} \Xcline{3-11}{1pt}
\multirow{4}{*}{\cite{DIV2K}} &                            & Original                                                        & \textcolor{red}{29.2358} & \textcolor{blue}{0.8251} & \textcolor{red}{0.2928} & 3.6466 & 2.9568 & \textcolor{red}{0.9907} & 1.9200 & 1.0836 \\ \cline{3-11} 
                       &                            & +SSIM                                                           & \textcolor{blue}{28.4237} & \textcolor{red}{0.8309} & \textcolor{blue}{0.2931} & 3.6586 & 2.9998 & 0.9861 & 2.0532 & 1.0928 \\ \cline{3-11} 
                       &                            & +EPBAN                                                 & 28.1644 & 0.7937 & 0.3200 & \textcolor{blue}{3.6703} & \textcolor{blue}{3.3650} & \textcolor{blue}{0.9893} & \textcolor{blue}{2.2832} & \textcolor{blue}{1.1187} \\ \cline{3-11} 
                       &                            & \begin{tabular}[c]{@{}c@{}}+SSIM$\&$EPBAN\end{tabular} & 28.2809 & 0.8171 & 0.3060 & \textcolor{red}{3.9231} & \textcolor{red}{3.4397} & 0.9885 & \textcolor{red}{2.5365} & \textcolor{red}{1.1433} \\ \Xhline{1pt}
                       
\multirow{4}{*}{\cite{B100}}  & \multirow{8}{*}{~\cite{linf}} & Original                                                        & \textcolor{blue}{25.0333} & 0.6646 & \textcolor{red}{0.2595} & \textcolor{blue}{4.2130} & 3.0861 & \textcolor{blue}{0.9818} & 2.7127 & 1.1549 \\ \cline{3-11} 
                       &                            & +SSIM                                                          & \textcolor{red}{25.9462} & \textcolor{red}{0.7154} & 0.3325 & 3.8825 & 2.9637 & \textcolor{red}{0.9824} & 2.3274 & 1.1169 \\ \cline{3-11} 
                       &                            & +EPBAN                                                & 24.0959 & 0.6389 & \textcolor{blue}{0.2783} & 4.2050 & \textcolor{blue}{3.3246} & 0.9816 & \textcolor{blue}{2.7503} & \textcolor{blue}{1.1585} \\ \cline{3-11} 
                       &                            & \begin{tabular}[c]{@{}c@{}}+SSIM$\&$EPBAN\end{tabular} & 24.7954 & \textcolor{blue}{0.6926} & 0.3122 & \textcolor{red}{4.2312} & \textcolor{red}{3.3978} & 0.9769 & \textcolor{red}{2.9046} & \textcolor{red}{1.1697} \\ \Xcline{1-1}{1pt} \Xcline{3-11}{1pt} 
\multirow{4}{*}{\cite{DIV2K}} &                            & Original                                                         & \textcolor{blue}{27.6269} & 0.7757 & \textcolor{red}{0.2172} & 3.8149 & 3.1076 & \textcolor{blue}{0.9877} & 2.4035 & 1.1293 \\ \cline{3-11} 
                       &                            & +SSIM                                                           & \textcolor{red}{28.6114} & \textcolor{red}{0.8176} & 0.2904 & 3.6578 & 3.0345 & \textcolor{red}{0.9881} & 2.0426 & 1.0936 \\ \cline{3-11} 
                       &                            & +EPBAN                                                 & 26.7167 & 0.7577 & \textcolor{blue}{0.2380} & \textcolor{red}{3.8275} & \textcolor{red}{3.3261} & 0.9876 & \textcolor{blue}{2.4411} & \textcolor{blue}{1.1330} \\ \cline{3-11} 
                       &                            & \begin{tabular}[c]{@{}c@{}}+SSIM$\&$EPBAN\end{tabular} & 27.3149 & \textcolor{blue}{0.7957} & 0.2722 & \textcolor{blue}{3.8261} & \textcolor{blue}{3.2854} & 0.9849 & \textcolor{red}{2.5422} & \textcolor{red}{1.1406} \\ \Xhline{1pt}

\end{tabular}
\label{t1}
\end{table*}
\vspace{-1em}
\section{Proposed Method}
\label{sec:2}
Our framework consists of two main components: an Efficient Perceptual Bi-directional Attention Network (Efficient-PBAN) trained as a perceptual quality metric, and its integration into SR optimization as a differentiable perceptual loss. The overall pipeline is illustrated in Fig.~1.

\subsection{The proposed Efficient-PBAN}
\label{section21}
\textbf{Feature Extraction.}  
Given an SR-HR image pair $\{x_{SR}, x_{HR}\}$, both inputs are processed by a ResNet stem and the first residual block (Layer1) with shared parameters. Beyond Layer1, the branches are separated to capture the distinct statistics of SR and HR images.

\textbf{PBA$^{+}$ Block.}  
For each input feature pair $(f^{SR}, f^{HR}) \in \mathbb{R}^{B \times C \times H \times W}$, query, key, and value representations are first obtained as:
\begin{equation}
\small
Q^{\gamma}, K^{\gamma}, V^{\gamma} = f_{\theta}(f^{\gamma}), \quad \gamma \in \{SR, HR\}.
\end{equation}
\begin{figure*}[t]
  \centering
  \includegraphics[width=0.9\textwidth]{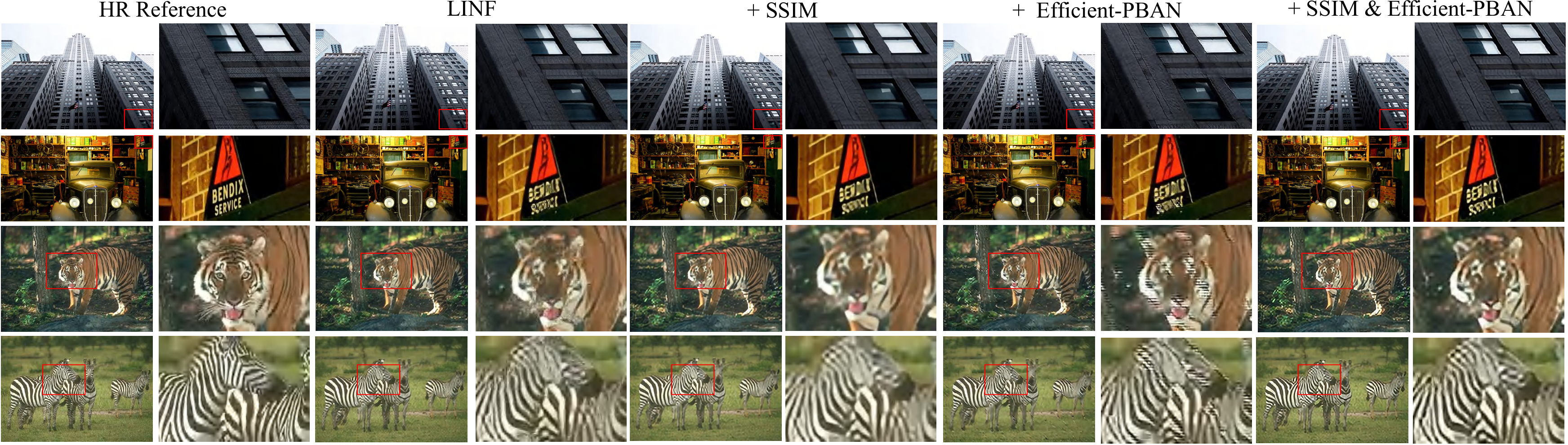}
  \caption{Visualizations of the original SR images and the corresponding optimization results, where the regions highlighted by red boxes are enlarged for clearer illustration.}
  \label{fig:visualization}
\end{figure*}

Attention is applied along both the height and width dimensions. For H-axis (column-wise) attention:
\begin{align}
\small
S^{H}_{HR \rightarrow SR} &= (Q^{H,SR})^{\top} K^{H,HR}, \\
A^{H}_{HR \rightarrow SR} &= \text{Softmax}\!\left(\frac{S^{H}_{HR \rightarrow SR}}
{\text{Std}(S^{H}_{HR \rightarrow SR})+\epsilon}\right), \\
O^{H}_{HR \rightarrow SR} &= V^{H,SR} A^{H}_{HR \rightarrow SR}.
\end{align}

Similarly, for W-axis (row-wise) attention:
\begin{align}
\small
S^{W}_{HR \rightarrow SR} &= (Q^{W,SR})^{\top} K^{W,HR}, \\
A^{W}_{HR \rightarrow SR} &= \text{Softmax}\!\left(\frac{S^{W}_{HR \rightarrow SR}}
{\text{Std}(S^{W}_{HR \rightarrow SR})+\epsilon}\right), \\
O^{W}_{HR \rightarrow SR} &= V^{W,SR} A^{W}_{HR \rightarrow SR}.
\end{align}

The final output features are the average of height- and width-axis attention:
\begin{equation}
\small
O_{HR \rightarrow SR} = \tfrac{1}{2}(O^{H}_{HR \rightarrow SR} + O^{W}_{HR \rightarrow SR}),
\end{equation}
with symmetric computation for $O_{SR \rightarrow HR}$. These outputs are fused with a SubEC~\cite{li2025pban} module to obtain enhanced bi-directional features as follows:
\begin{equation}
\small
\begin{aligned}
&I = \{O_{HR \rightarrow SR}, O_{SR \rightarrow HR}\},\\
&\{\hat{O}_{HR \rightarrow SR}, \hat{O}_{SR \rightarrow HR}\}\\
&=\{f^{SR},f^{HR}\}+I\times f_{Sub-Channel}(I)\times f_{Sub-Pixel}(I),
\end{aligned}
\end{equation}
where $f_{Sub-Channel}(\cdot)$ and $f_{Sub-Pixel}(\cdot)$ employ learnable upsampling and shuffle operation to extract sub-pixel and sub-channel cues at spatial and channel dimensions, respectively.

\textbf{Quality Prediction Module.} 
The enhanced features are passed to a prediction head with global pooling and fully-connected layers to regress a perceptual score $\hat{q}$ as follows:
\begin{equation}
\small
\begin{aligned}
&f_{Quality}(\cdot) = D(ReLU(L(D(ReLU(L(F(P(\cdot)))))))),\\
&\{O^{'}_{HR \rightarrow SR},O^{'}_{SR \rightarrow HR}\}=f_{Quality}(\{\hat{O}_{HR \rightarrow SR}, \hat{O}_{SR \rightarrow HR}\}),\\
&\hat{q} = L(L(Concate(O^{'}_{HR \rightarrow SR},O^{'}_{SR \rightarrow HR}))),
\end{aligned}
\end{equation}
where $D, L, F, P$ refer to the Dropout, Linear, Flatten, and Average pooling operations, respectively.

We then perform the training with an $L_{2}$ regression loss:
\begin{equation}
\small
\mathcal{L}_{Q} = \| \hat{q} - q \|^{2},
\end{equation}
where $q$ is the ground-truth human opinion score.
\begin{table*}[!ht]
    \centering
    \small
    \caption{Ablation study on the weighting ratio between the distortion-oriented loss $\mathcal{L}_{D}$ (denoted as $\alpha$) and the perceptual loss $\mathcal{L}_{P}$ (denoted as $\beta$) for CAMixerSR on the B100 dataset. The best performances are highlighted in bold.}
    \begin{tabular}{c|cccccccc}
    \hline
        $\beta$/$\alpha$ & PSNR $\uparrow$ & SSIM $\uparrow$& LPIPS $\downarrow$& PFIQA$\uparrow$ & Efficient-PBAN$\uparrow$ & SF$\uparrow$ & SN$\uparrow$ & SFSN $\uparrow$ \\ \hline
        1/9 & 26.1260 & \textbf{0.7195} & 0.3315 & 3.8717 & 2.6703 & 0.9848 & 2.1308 & 1.0994  \\ \hline
        2/8 & 26.1273 & \textbf{0.7195} & 0.3314 & 3.8716 & 2.6707 & 0.9848 & 2.1302 & 1.0993  \\ \hline
        3/7 & 26.1292 & 0.7194 & 0.3314 & 3.8710 & 2.6712 & 0.9848 & 2.1292 & 1.0992  \\ \hline
        4/6 & \textbf{26.1334} & 0.7192 & \textbf{0.3313} & 3.8692 & 2.6725 & \textbf{0.9849} & 2.1269 & 1.0991  \\ \hline
        5/5 & 25.6741 & 0.7110 & 0.3560 & 3.9348 & 3.4025 & 0.9843 & 2.4084 & 1.1268  \\ \hline
        6/4 & 20.2765 & 0.3358 & 0.5914 & \textbf{4.9893} & 4.3056 & 0.9699 & \textbf{3.4414} & \textbf{1.2171}  \\ \hline
        7/3 & 20.2822 & 0.3441 & 0.5846 & 4.9660 & 4.3398 & 0.9675 & 3.4389 & 1.2146  \\ \hline
        8/2 & 20.2670 & 0.3472 & 0.5821 & 4.3576 & \textbf{4.9588} & 0.9679 & 3.4390 & 1.2150  \\ \hline
        9/1 & 20.2551 & 0.3489 & 0.5806 & 4.9528 & 4.3687 & 0.9675 & 3.4389 & 1.2146 \\ \hline
    \end{tabular}
    \label{t2}
    \vspace{-1.3em}
\end{table*}
\vspace{-1em}
\subsection{Perceptual Optimization}
\label{sec:2.2}
After pre-training Efficient-PBAN as a quality predictor, it is integrated into SR training as a perceptual loss. Given an LR input, the SR model generates a reconstruction $x_{SR}$, which is compared with the HR reference $x_{HR}$ by Efficient-PBAN. To reduce the window effect caused by patch-based SR methods, our method combines distortion-oriented and perceptual losses:
\begin{equation}
\small
\begin{aligned}
&\ \mathcal{L}_{D} = -Q(x_{SR}, x_{HR},\theta_{SSIM}),\\
&\ \mathcal{L}_{P} = - Q(x_{SR}, x_{HR}; \theta_{Efficient-PBAN}),\\
&\ \mathcal{L} = \alpha\times\frac{\mathcal{L}_{D}}{\mathcal{L}_{D}+\mathcal{L}_{P}}+\beta\times\frac{\mathcal{L}_{P}}{\mathcal{L}_{D}+\mathcal{L}_{P}}.
\end{aligned}
\end{equation}
where $Q(\cdot)$ is the differentiable quality prediction output, and $\alpha,\beta$ are weighting ratios of the two losses. By minimizing $\mathcal{L}$, the SR network is guided to maximize perceptual quality scores aligned with human visual preferences.
\vspace{-1em}
\section{Experimental Results and Analysis}
\subsection{Databases}
Following the protocols of \cite{wang2024camixersr,linf}, we train on the DIV2K~\cite{DIV2K} dataset and validate on both the B100~\cite{B100} dataset and the validation set of DIV2K. B100 consists of 100 natural images with diverse content and structures. DIV2K provides 1,000 2K-resolution images with rich textures and semantics, serving as standard benchmarks for image super-resolution.

In addition, we construct the latest SR quality database to pre-train Efficient-PBAN. The database comprises 720 SR images at approximately 2K resolution, generated from 19 high-resolution reference images selected from the DIV2K dataset by maximizing the envelope area in the Spatial Information-Colorfulness space to ensure content diversity. SR images are produced at four upscaling factors (×2, ×3, ×4, and ×8), with 576 samples for training, 72 for validation, and 72 for testing. The dataset covers a broad range of state-of-the-art SR methods, including 4 GAN-based, 5 diffusion-based, 4 transformer-based, 1 flow-based, and 1 CNN-based approach, ensuring strong alignment between the learned metric and subjective perceptual judgments. Subjective quality scores are collected through a single-stimulus experiment conducted under ITU-R BT.500-14 conditions, involving 23 participants (aged 18–50) from five regions with normal or corrected vision. This database has also been adopted in a successful Grand Challenge, ISRGen-QA~\cite{li2025vquala2025challengeimage}.
\vspace{-1em}
\subsection{Experimental Setup}
We evaluate perceptual optimization on two representative SR baselines, CAMixerSR~\cite{wang2024camixersr} and LINF~\cite{linf}, under four settings: (1) Original; (2) +SSIM; (3) +Efficient-PBAN; and (4) +SSIM \& Efficient-PBAN ($\alpha=\beta=0.5$). Performance is assessed using distortion metrics (PSNR, SSIM~\cite{SSIM}), perceptual metrics (LPIPS~\cite{lpips}, PFIQA~\cite{pfiqa}), our Efficient-PBAN score, and SR-specific indices (SF, SN, SFSN~\cite{sfsn}).
\vspace{-0.5em}
\subsection{Experimental Results}
Quantitative results on B100 and DIV2K are summarized in Table~\ref{t1}. First, Efficient-PBAN optimization consistently improves perceptual quality, achieving higher PFIQA and Efficient-PBAN scores than both Original and SSIM-based models. Second, although perceptual optimization slightly degrades PSNR/SSIM, perceptual metrics (LPIPS, PFIQA) are substantially improved, indicating better alignment with human perception. Third, joint optimization with SSIM and Efficient-PBAN achieves a favorable trade-off.

Qualitative comparisons in Fig.~\ref{fig:visualization} further show that Efficient-PBAN-guided models recover finer textures and sharper edges, whereas Original and SSIM-based models tend to produce over-smoothed details, confirming the effectiveness of metric-driven perceptual optimization. However, since the latest SR methods rely on patch-based learning, the standalone Efficient-PBAN lacks an overall fidelity constraint, which leads to the emergence of window artifacts.

Table~\ref{t2} reports an ablation study on the loss weighting between distortion loss $\mathcal{L}_D$ ($\alpha$) and perceptual loss $\mathcal{L}_P$ ($\beta$). When $\beta/\alpha$ is small (1/9–4/6), SF and PSNR remain high, indicating strong structural fidelity but limited perceptual naturalness. Increasing $\beta/\alpha$ emphasizes perceptual quality, reducing SF and PSNR while significantly improving SN, which peaks at $\beta/\alpha=6/4$. This trend clearly illustrates the trade-off between reconstruction fidelity and visual realism.

A subjective test study was further conducted following~\cite{li2025vquala2025challengeimage}, with MOS results consistent with Table~\ref{t1}. For CAMixerSR, the ranking is EPBAN \& SSIM (3.25/5) $>$ EPBAN (3.00/5) $>$ SSIM (2.92/5) $>$ Original (2.83/5). For LINF, EPBAN \& SSIM (2.75/5) $>$ SSIM (2.50/5) $>$ Original (2.42/5) $>$ EPBAN (1.92/5). These results confirm that joint optimization yields the best perceptual quality, while Efficient-PBAN or SSIM alone provides moderate improvements over the original SR outputs.
  \vspace{-1em}
\section{Conclusion}
\label{sec:4}
In this work, we proposed Efficient-PBAN, a perceptual quality optimization framework for image super-resolution. By constructing the latest SR quality dataset and training a bi-directional attention-based quality predictor, we enabled closed-loop optimization directly aligned with human visual perception. Extensive experiments demonstrate that our approach improves perceptual quality over state-of-the-art baselines, validated by both qualitative and quantitative aspects. In future work, we plan to extend our framework to more complex generative SR paradigms, such as more diffusion-based models, and to further expand the SR quality database for broader coverage.

\bibliographystyle{IEEEbib}
\bibliography{refs}

\end{document}